\documentclass[twocolumn,showpacs,preprintnumbers,amsmath,amssymb]{revtex4-1}
\usepackage[utf8]{inputenc}
\usepackage{graphicx}
\usepackage{dcolumn}
\usepackage{float}
\usepackage{color}
\usepackage{bm}
\def\MM{{\Bbb M}}

\begin{document}

\title{\color{blue} A neutrino mixing model based on an $A_4\times Z_3\times Z_4$ 
flavour symmetry}

\author{Nguyen Anh Ky$^{1-3}$}
\def\thefootnote{a}
\email{anhky@iop.vast.vn}
\email{anh.ky.nguyen@cern.ch}
\author{Phi Quang V$\breve{\mbox{a}}$n$^{2}$}
\email{pqvan@iop.vast.vn}
\author{Nguyen Thi H$\grave{\hat{\mbox{o}}}$ng V$\hat{\mbox{a}}$n$^{2}$}
\email{nhvan@iop.vast.vn}
\affiliation{\vspace*{2mm}
$^1$\textit{Duy Tan university},\\
\textit{K7/25 Quang Trung street, Hai Chau, Da Nang, Viet Nam.}\vspace*{3mm}\\
$^2$\textit{Mathematical and high energy physics group},\\ 
\textit{Institute of physics},\\ 
\textit{Vietnam academy of science and technology (VAST)},\\ 
\textit{10 Dao Tan, Ba Dinh, Hanoi, Viet Nam.} 
\vspace*{3mm}\\
$^3$\textit{Laboratory of high energy physics and cosmology},\\ 
\textit{Faculty of physics}, \\
\textit{VNU university of science},\\ 
%{\it Vietnam national university},\\ 
\textit{334 Nguyen Trai, Thanh Xuan, Hanoi, Viet Nam.}\\
}

\date{\today}

\begin{abstract}
A model of a neutrino mixing with an $A_4\times Z_3\times Z_4$ flavour symmetry is suggested. 
In addition to the standard model fields, the present model contains six new fields which 
transform under different representations of $A_4\times Z_3\times Z_4$. The model is constructed 
to slightly deviate from a tri-bi-maximal model in agreement with the current experimental data, 
thus, all analysis can be done in the base of the perturbation method. Within this model, as an 
application, a relation between the mixing angles ($\theta_{12}, \theta_{23}, \theta_{13}$) 
and the Dirac CP-violation phase ($\delta_{CP}$) is established. This relation allows a 
prediction of $\delta_{CP}$ and the Jarlskog parameter ($J_{CP}$). The predicted value 
$\delta_{CP}$ is in the 1$\sigma$ region of the global fit for both the normal- and inverse 
neutrino mass ordering and gives $J_{CP}$ to be within the bound $|J_{CP}|\leq 0.04$. For an 
illustration, the model is checked numerically and gives values of the neutrino masses (of 
the order of 0.1 eV)  and the mixing angle $\theta_{13}$ (about $9^\circ$) very close to the 
current experimental data. 
\end{abstract}

\maketitle

\section{\label{sec:level1}Introduction}
  
~~~ 
After the discovery of the Higgs boson, called also Brout-Englert-Higgs (BEH) boson, 
\cite{Aad:2012tfa,Chatrchyan:2012ufa} by the LHC  collaborations ATLAS and CMS (for 
a review, see, for example, 
\cite{Ky:2015eka}), 
the particle content of the standard model (SM) seems to be completely confirmed by 
the experiment. The SM is an excellent model of elementary particles and their 
interactions as it can explain and predict many phenomena, at least until the energy 
scale around the top quark mass. However, there are open problems which cannot be solved 
within the SM and thus call for modifying or extending the latter. The problem of 
neutrino masses and mixings 
\cite{Fukuda:1998tw,Fukuda:1998ub,Fukuda:1998mi,Ahmad:2001an,Ahmad:2002jz,Ahmad:2002ka} 
is among 
such problems beyond the SM.
This problem is important for not only particle physics but also nuclear physics, 
astrophysics and cosmology, therefore, it has attracted much interest \cite{Bilen,Carlo,
Mohapatra:1998rq,Lesgou,Adhikari:2016bei}. The neutrino mixing means that the flavour neutrinos (flavour 
eigenstates of neutrinos) are superpositions of massive neutrinos (mass eigenstates 
of neutrinos) encoded in the so-called Pontecorvo-Maki-Nakagawa-Sakata 
(PMNS) matrix in terms of mixing angles $\theta_{ij}$ and a given number of phases,   
while in the SM the neutrinos are massless and not mixing. One of the ways trying to 
explain this phenomenon is to add a flavour symmetry to the gauge symmetry 
$SU(3)_c\otimes SU(2)_L\otimes U(1)_Y$ 
of the SM (see \cite{Ishimori:2010au,Altarelli:2010gt} for a review). A popular flavour 
symmetry intensively investigated in literature is that described by the group $A_4$ (see, 
for instance, 
\cite{Altarelli:2010gt, Hirsch:2007kh, Altarelli:2005yx, Altarelli:2009kr,Parattu:2010cy,
King:2011ab,Altarelli:2012bn, 
Altarelli:2012ss,Ferreira:2013oga}) allowing to obtain a tribi-maximal (TBM) neutrino mixing 
corresponding to the mixing angles 
$\theta_{12}\approx 35.26^{\circ}$ ($\sin^2\theta_{12}=1/3$), $\theta_{13}=0^{\circ}$ and 
$\theta_{23}=45^{\circ}$ (see \cite{Harrison:2002er}). The recent experimental data such as that 
from T2K \cite{Abe:2014ugx,Abe:2013hdq}, RENO \cite{Ahn:2012nd}, DOUBLE-CHOOZ \cite{Abe:2012tg}, 
DAYA-BAY \cite{An:2013zwz,Hu:2014tia} showing a non-zero mixing angle $\theta_{13}$ and a possible 
non-zero Dirac CP-violation (CPV) phase $\delta_{CP}$, rejects, however, the TBM scheme 
\cite{Capozzi:2013csa,Agashe:2014kda}. There have been many attempts to explain these experimental 
phenomena. In particular, for this purpose, various models with a discrete flavour symmetry 
\cite{Ishimori:2010au,Smirnov:2013uba,Hernandez:2013vya,Hernandez:2012ra}, including an $A_4$ flavour 
symmetry, have been suggested 
\cite{Ishimori:2010au,Altarelli:2010gt, Hirsch:2007kh,  Altarelli:2005yx, 
Altarelli:2009kr,Parattu:2010cy,King:2011ab,Altarelli:2012bn,
Altarelli:2012ss,Ferreira:2013oga,Ma:2004zv,Barry:2010zk,Ahn:2012tv,Frampton:2002yf,
Ma:2001dn,Babu:2002dz,Ishimori:2012fg,Ma:2012ez,Dinh:2015tna,Dinh:2016tuk,Hung:2015nva}. \\

In general, the models, based on $A_4$ flavour symmetry, have extended lepton and scalar sectors 
containing new fields in additions to the SM ones which now may have an $A_4$ symmetry structure. 
Therefore, besides undergoing the SM symmetry, these fields may also transform under $A_4$. 
At the beginning, the $A_4$-based models were build to describe a TBM neutrino mixing 
(see, for example, \cite{Altarelli:2005yx}) but later many attempts, such as those in 
\cite{Ishimori:2010au,Altarelli:2010gt, Hirsch:2007kh, 
Altarelli:2009kr,Parattu:2010cy,King:2011ab,Altarelli:2012bn,
Altarelli:2012ss,Ferreira:2013oga,Barry:2010zk,Ahn:2012tv,Ishimori:2012fg,Ma:2012ez,
Dinh:2015tna,Dinh:2016tuk}, 
to find a model fitting the non-TBM phenomenology, have been made. 
On 
these models, however, are often imposed some assumptions, for example, the vacuum expectation 
values (VEV's) of some of the fields, especially those generating neutrino masses, have a 
particular alignment \cite{Frampton:2002yf,Ma:2001dn,Babu:2002dz,Ishimori:2012fg,Ma:2012ez}. 
These assumptions may lead to a simpler diagonalization of a mass matrix but restrict the 
generality of the model.
Since, according to the current experimental data, the discrepancy of a phenomenological  
model from a TBM model (i.e., a model in which the neutrino mixing matrix has a TBM form 
\cite{Harrison:2002er}) is quite small, we can think about a perturbation approach to 
building a new, realistic, model 
\cite{Dinh:2015tna}. \\

%%%%%%%%%%%%%%%%%%%%%%%%%%%%%%%%%%%%%

   The perturbative approach has been used by several authors (see for example, 
\cite{Brahmachari:2012cq, Brahmachari:2014npa}) but their methods mostly are model-independent, 
that is, no model realizing 
the experimentally established neutrino mixing has been shown. On the other hand, most 
of the $A_4$-based models are analyzed in a non-perturbative way. There are a few cases such 
as \cite{Honda:2008rs} 
where the perturbative method is applied but their approach is different from ours and their 
analysis, sometimes, is not precise (for example, the conditions imposed in section IV of 
\cite{Honda:2008rs} are not always possible). Besides that, in many works done so far, 
the neutrino mixing has been investigated with a less general
vacuum structure of scalar fields.\\  
                              
  In this paper we will introduce an $A_4$ flavour symmetric standard model, 
which can generate a neutrino mixing, deviating from the TBM scheme slightly, as requested and 
explained above. Since the deviation is small we can use a perturbation method in elaborating 
such a non-TBM neutrino mixing model. The model field content is that of the SM extended with 
six new fields, all are $SU(2)_L$ singlets (iso-singlets),  transforming under different 
representations of $A_4$: an $A_4$ triplet fermion $N$, two $A_4$ triplet scalars $\varphi_E$ 
and $\varphi_N$, and three $A_4$ singlet scalars $\xi$, $\xi^{'}$ and $\xi^{''}$. In order to 
exclude unwanted interactions two additional symmetries, namely $Z_3$ and $Z_4$, are imposed, 
and, as a result, the model is based on an $A_4\times Z_3\times Z_4$ flavour symmetry times the 
SM symmetry  
(see Table \ref{a4} for more detailed group transformations of the lepton and scalar fields 
in this extended model). For generality we consider a scalar sector containing all possible 
representations of $A_4$. The presence of the fields $\xi^{'}$ and $\xi^{''}$ guarantees a 
nontrivial mass matrix of the charged leptons, otherwise, the latter would become massless. 
The corresponding neutrino mass matrix can be developed perturbatively around a neutrino mass 
matrix diagonalizable by a TBM mixing matrix. As a consequence, a relation between the Dirac 
CPV phase $\delta_{CP}$ and the mixing angles $\theta_{ij}$, $i,j=1,2,3$ (for a three-neutrino 
mixing model) are established. Based on the experimental data of the mixing angles, this relation 
allows us to determine $\delta_{CP}$ numerically in both normal ordering (NO) and inverse ordering 
(IO). It is very important as the existence of a Dirac CPV phase indicates a difference between 
the probabilities $P(\nu_l\rightarrow \nu_{l'})$ and $P(\bar{\nu}_l\rightarrow \bar{\nu}_{l'})$, 
$l\neq l'$, of the neutrino- and antineutrino transitions (oscillations) in vacuum 
$\nu_l\rightarrow \nu_{l'}$ and $\bar{\nu}_l\rightarrow \bar{\nu}_{l'}$, 
respectively, thus, a CP violation in the neutrino subsector of the lepton sector. 
We should note that for a three-neutrino mixing model, as considered in this paper, the mixing matrix 
in general has one Dirac- and two Majorana CPV phases \cite{Bilenky:1980cx} (for a more general, 
$n$-neutrino mixing, case, see \cite{Schechter:1980gr,Doi:1980yb}). Since the Majorana CPV phases do 
not effect these transition probabilities they are not a subject of a detailed analysis  here. \\

In the framework of the suggested model and the perturbation method our approach allows us to 
obtain $\delta_{CP}$ within 
the $1\sigma$ region of the best fit value \cite{Agashe:2014kda}. This approach is different but 
our result is 
quite consistent with that obtained by other authors (see, for example, 
\cite{Girardi:2015zva,Girardi:2014faa,Petcov:2014laa,Petcov:2013poa,Kang:2015xfa,
Gonzalez-Garcia:2014bfa,Kang:2014mka} and references therein). Further, knowing 
$\delta_{CP}$ we can determine the Jarlskog parameter ($J_{CP}$) measuring a CP violation. The 
determination of $\delta_{CP}$ and $J_{CP}$ represents an application of the present model and,  
in this way, verifies the latter (of course, it is not a complete verification). 
A numerical test of the model gives values of the neutrino masses,  
the mixing angle $\theta_{13}$ and the Dirac CP-violation phase consistent with the 
current experimental results.\\
 
This paper has the following plan. A brief introduction to the representations of $A_4$ and 
their application to building an extended standard model will be made in the next section. 
Neutrino masses and mixing within this model are considered in Sect. 3 via a perturbation method. 
Sect. 4 is devoted to the investigation of Dirac CPV phase and Jarlskog parameter. The last section 
is designed for some discussions and conclusions.

\section{Extended standard model with an $A_4\times Z_3\times Z_4$ flavour symmetry}

~~~ Here, we will deal with an extended SM acquiring an additional $A_4$ flavour symmetry. 
An extra $Z_3\times Z_4$ symmetry is also introduced to constrain the model not to deviate 
too much from the SM. As mentioned above the flavour symmetry, in particular, that based on 
the group $A_4$, has attracted much interest during last about ten years 
(see \cite{Ishimori:2010au,Altarelli:2010gt} for a review). Let us first summarize here 
representations of $A_4$ \cite{Ishimori:2010au,Altarelli:2012bn, King:2013eh} and then 
review briefly the model which will be considered. 

\subsection{Summary of representations of $A_4$}
\vspace{1mm}

~~~ The group $A_4$ is a group of even-permutations on four objects and thus it has 12 elements ($12=4!/2$). 
This group is also called the tetrahedral group as it can describe the orientation-preserving symmetry of 
a regular tetrahedron. It can be generated by two basic permutations $S$ and $T$ having properties 
\begin{equation}
S^2=T^3=(ST)^3=1.
\end{equation}
The group representations are relatively simple and include three one-dimensional unitary representations 
1, $1^{'}$ and $1^{''}$ with the generators $S$ and $T$ given, respectively, as follows
\begin{subequations}
\begin{equation}
1: \hspace{0.2cm} S=1, \hspace{0.3cm} T=1,
\end{equation}
\begin{equation}
1^{'}:\hspace{0.2cm} S=1, \hspace{0.3cm} T=e^{i2 \pi/3} \equiv \omega, 
\end{equation}
\begin{equation}
1^{''}: \hspace{0.2cm} S=1, \hspace{0.3cm} T=e^{i4 \pi/3} \equiv \omega^2,
\end{equation}
\end{subequations}
and a three-dimensional unitary representation with the generators
\begin{equation}
T = \left(
\begin{array}{ccc}
 1 & 0 & 0 \\
 0 & \omega^2 & 0 \\
 0 & 0 & \omega
\end{array}
\right),
\hspace{0.3cm}
S=\frac{1}{3} \left(
\begin{array}{ccc}
 -1 & 2 & 2 \\
 2 & -1 & 2 \\
 2 & 2 & -1
\end{array}
\right).\vspace*{.1mm}
\end{equation}
Here we use the three-dimensional representation where the generator $T$ has a diagonal 
form \cite{Altarelli:2005yx}. The reason of choosing this representation is that the latter 
ensures the diagonal mass matrix of the charged leptons (see the next section).\\

Representation theory and applications of a group often require to know a  multiplication and 
decomposition rule of a product of its (irreducible) representations. In the case of $A_4$ these 
rules read 
\begin{subequations}
\begin{equation}
1\times 1 = 1,
\end{equation}
\begin{equation}
1^{'} \times 1^{''} = 1,
\end{equation}
\begin{equation}
1^{''} \times 1^{'} = 1,
\end{equation}
\begin{equation}
1^{'} \times 1^{'} = 1^{''},
\end{equation}
\begin{equation}
1^{''} \times 1^{''} = 1^{'},
\end{equation}
\begin{equation}
3 \times 3 = 1+1^{'}+1^{''} +3_s+3_{a}.
\label{33prod}
\end{equation}
\end{subequations}
While the first five rules are trivial, let us give more explicit expressions for the multiplication 
and decomposition rule for a product \eqref{33prod} between two triplets, say $3_a \sim (a_1,a_2,a_3)$ 
and $3_b \sim (b_1,b_2,b_3)$. This direct product can be decomposed into three singlets and two triplets 
as follows
\begin{widetext}
\begin{subequations}
\begin{equation}
1 = a_1b_1+a_2b_3+a_3b_2,
\end{equation}
\begin{equation}
1^{'} = a_3b_3+a_1b_2+a_2b_1,
\end{equation}
\begin{equation}
1^{''} = a_2b_2+ a_1b_3+a_3b_1,
\end{equation}
\begin{equation}
3_s \sim  \frac{1}{3}(2a_1b_1-a_2b_3-a_3b_2, 2a_3b_3-a_1b_2-a_2b_1, 2a_2b_2-a_1b_3-a_3b_1),
\end{equation}
\begin{equation}
3_{a} \sim \frac{1}{3}(a_2b_3-a_3b_2, a_1b_2-a_2b_1, a_1b_3-a_3b_1).
\end{equation}
\end{subequations}
\end{widetext}
\vspace*{2mm}

The above-given information will be used for the construction of a Lagrangian, as the one in 
\eqref{lagran}, of a model with an $A_4$ symmetry.

\subsection{The model}
\vspace{1mm}

~~~ Compared with the SM, the model studied here contains an extended lepton- and scalar 
sector (the quark sector is not considered here yet). The lepton sector includes an 
$A_4$-triplet $N$ (its components are referred to as right-handed neutrinos), which is an 
iso-singlet, in addition to the SM leptons among which the left-handed lepton iso-doublets 
$\ell_L$, $\ell=\tilde{e}, \tilde{\mu},\tilde{\tau}$,  all together form an $A_4$-triplet, 
while the right-handed lepton iso-singlets $\tilde{e}_R$, $\tilde{\mu}_R$ and 
$\tilde{\tau}_R$ transform as $A_4$-singlets $1$, $1^{'}$ and $1^{''}$, respectively. In 
general, the basis $\ell=\tilde{e}, \tilde{\mu},\tilde{\tau}$, in which the charged lepton 
mass matrix may not be diagonal, are different from the standard basis of the mass states 
$l=e, \mu, \tau$. Besides 
the original SM Higgs $\phi_h$ which is an $A_4$-singlet, the scalar sector 
of the model has five additional iso-singlet fields: two  $A_4$-triplets $\varphi_{E}$ 
and $\varphi_{N}$, and three $A_4$-singlets $\xi$, $\xi^{'}$, $\xi^{''}$. 
Our choice of the model field content, thus,  
covers all irreducible representations of $A_4$. To keep 
maximally the SM interaction structure (as many its consequences have been experimentally 
verified very well) an additional $Z_3\times Z_4$ symmetry is introduced. The transformation 
rules under $SU(2)_L$, $A_4$, $Z_3$ and $Z_4$ of the leptons and the scalars in this model 
are summarized in Table \ref{a4}. Let us look at a closer distance the scalar- and the 
lepton sector.
\begin{table}[H]
%~\\[4mm]
\begin{center}
\begin{tabular}{|l||c|c|c|c|c|c|c|c|c|c|c|}
\hline
\vspace*{-2.5mm}
&&&&&&&&&&&\\
  & $\ell_L$ & $\tilde{e}_R$ & $\tilde{\mu}_R$ & $\tilde{\tau}_R$ & $\phi_h$ & $N$ & $\varphi_E$ &$\varphi_N$ 
& $\xi$ & $\xi^{'}$ & $\xi^{''}$  \\ 
  \vspace*{-2.5mm}
  &&&&&&&&&&&\\
   \hline
 \vspace*{-2.5mm}
 &&&&&&&&&&&\\
 $SU(2)_L$\vspace*{-2.5mm} & 2 & 1 & 1 & 1 & 2 & 1 & 1 & 1 & 1 & 1 & 1   \\
 %\vspace*{-2mm}
 &&&&&&&&&&&\\
 \hline 
   \vspace*{-2.5mm}
  &&&&&&&&&&&\\
 $A_4$ & 3 & 1   & $1^{'}$ & $1^{''}$ & 1 &  3   &  3 & 3  &  1  & $1^{'}$   & $1^{''}$     \\  
 \vspace*{-2.5mm}
 &&&&&&&&&&&\\
\hline 
 \vspace*{-2.5mm}
 &&&&&&&&&&&\\
 $Z_3$\vspace*{-2.5mm} & $\omega^2$ & 1 & 1 & 1 & $\omega^2$ & $\omega$ & 1 & $\omega$ 
& $\omega$ & $\omega$ & $\omega^2$   \\
 %\vspace*{-2mm}
 &&&&&&&&&&&\\
 \hline
  \vspace*{-2.5mm}
 &&&&&&&&&&&\\
 $Z_4$\vspace*{-2.5mm} & i & 1 & 1 & 1 & 1 & i & i & -1 
& -1 & i & i   \\
 %\vspace*{-2mm}
 &&&&&&&&&&&\\
  \hline 
\end{tabular}
\caption{Lepton- and scalar sectors of the model and their group transformations, where 
$\omega^k=e^{2k\pi/3}$, $k=0,1,2$.}
\label{a4}
\end{center}
\end{table}
\subsubsection{Scalar sector}
~~~ The scalar potential has the form 
\begin{widetext}
\begin{align}
V (\phi_h,\varphi_E, \varphi_N, \xi,\xi^{'},\xi^{''} )=V_1 (\phi_h)+V_2 (\varphi_E, \xi^{'}, \xi^{''})  
+V_3(\varphi_N,\phi_h,\xi, \xi^{'},\xi^{''}) +V_4(\xi, \phi_h),
\end{align}
with  
\begin{equation}
V_1 (\phi_h)=\mu^2 (\phi_h^{\dagger} \phi_h) + \lambda_0 (\phi_h^{\dagger} \phi_h )^2,
\end{equation}
\begin{align}
V_2 (\varphi_E, \xi^{'}, \xi^{''})=&~\alpha_1 (\varphi_E \varphi_E)_{\mathbf{1}}(\varphi_E \varphi_E)_{\mathbf{1}}
+\alpha_2 (\varphi_E \varphi_E)_{\mathbf{1^{'}}}(\varphi_E \varphi_E)_{\mathbf{1^{''}}} \nonumber  \\
&+\alpha_3 (\varphi_E \varphi_E)_{\mathbf{3_s}}(\varphi_E \varphi_E)_{\mathbf{3_s}}  
+\alpha_4 (\varphi_E \varphi_E)_{\mathbf{3_a}}(\varphi_E \varphi_E)_{\mathbf{3_a}}  \nonumber  \\
&+\alpha_5 (\varphi_E \varphi_E)_{\mathbf{3_s}}(\varphi_E \varphi_E)_{\mathbf{3_a}}   
%
%\nonumber  \\&
%
+\left[ \frac{\alpha_6}{2} (\varphi_E \varphi_E)_{\mathbf{1}} (\xi^{'}\xi^{''})_{\mathbf{1}}+ h.c.\right],
%\frac{\alpha_6^{\ast}}{2} (\varphi_E \varphi_E)_{\mathbf{1}} (\xi^{' \dagger} \xi^{'' \dagger})_{\mathbf{1}},
\end{align}
\begin{align}
V_3(\varphi_N,\phi_h,\xi, \xi^{'},\xi^{''})=&~\mu^2 \left(\varphi_N^{\dagger} \varphi_N \right)_{\mathbf{1}}
+ \lambda_1 \left(\varphi_N^{\dagger} \varphi_N \right)_{\mathbf{1}}^2 
+ 2\lambda_2 \left(\varphi_N^{\dagger} \varphi_N \right)_{\mathbf{1^{'}}} 
\left(\varphi_N^{\dagger} \varphi_N \right)_{\mathbf{1^{''}}}   \nonumber  \\
&+ \lambda_3 \left(\varphi_N^{\dagger} \varphi_N \right)_{\mathbf{3_s}} 
\left(\varphi_N^{\dagger} \varphi_N \right)_{\mathbf{3_s}}   
+ \lambda_4 \left(\varphi_N^{\dagger} \varphi_N \right)_{\mathbf{3_a}} 
\left(\varphi_N^{\dagger} \varphi_N \right)_{\mathbf{3_a}} \nonumber  \\
&+ 2\lambda_5 \left(\varphi_N^{\dagger} \varphi_N \right)_{\mathbf{3_{s}}} 
\left(\varphi_N^{\dagger} \varphi_N \right)_{\mathbf{3_a}} \nonumber  \\
&+\gamma_1 \left(\varphi_N^{\dagger} \varphi_N \right)_{\mathbf{1}} \left( \xi^{\dagger} \xi \right)_{\mathbf{1}} 
+\gamma_2 \left(\varphi_N^{\dagger} \varphi_N \right)_{\mathbf{1^{''}}}  \left( \xi^{'' \dagger} \xi^{''} \right)_{\mathbf{1^{'}}} \nonumber  \\
&+\gamma_3 \left(\varphi_N^{\dagger} \varphi_N \right)_{\mathbf{1^{'}}} \left( \xi^{' \dagger} \xi^{'} \right)_{\mathbf{1^{''}}} 
+\gamma \left(\varphi_N^{\dagger} \varphi_N \right)_{\mathbf{1}} \left(\phi_h^{\dagger} \phi_h \right)_{\mathbf{1}},  
\label{VphiH}
\end{align}
and 
\begin{align}
V_4(\xi, \phi_h)= \eta_1^2 \left(\xi^{\dagger} \xi \right)_{\mathbf{1}}+\chi_1 \left(\xi^{\dagger} \xi \right)_{\mathbf{1}}^2
+\chi_2\left(\xi^{\dagger} \xi \right)_{\mathbf{1}} \left( \phi_h^{\dagger} \phi_h \right)_{\mathbf{1}}.
\end{align}
\end{widetext}
 Here, the coefficients $\lambda_2$ and $\lambda_5$ are multiplied by 2 just for further convenience. 
The additional $Z_3\times Z_4$ symmetry is introduced in order to avoid interactions 
between the scalar fields $\varphi_E$ and $\varphi_N$ which would be  
\begin{widetext}
\begin{align}
V_5(\varphi_E, \varphi_N)=&~
\rho_1 (\varphi_E \varphi_E )_{\mathbf{3_s}} 
(\varphi_N^{\dagger} \varphi_N )_{\mathbf{3_s}} +\rho_2 (\varphi_E \varphi_E )_{\mathbf{3_s}} 
(\varphi_N^{\dagger} \varphi_N )_{\mathbf{3_a}}    
%\nonumber  \\
+ \rho_3 (\varphi_E \varphi_E )_{\mathbf{3_a}} 
(\varphi_N^{\dagger} \varphi_N )_{\mathbf{3_s}} 
+\rho_4 (\varphi_E \varphi_E )_{\mathbf{3_a}} 
(\varphi_N^{\dagger} \varphi_N )_{\mathbf{3_a}}   
 \nonumber  \\
&+\rho_5 (\varphi_E \varphi_E )_{\mathbf{1}} 
(\varphi_N^{\dagger} \varphi_N )_{\mathbf{1}} +\rho_6 (\varphi_E \varphi_E )_{\mathbf{1^{'}}} 
(\varphi_N^{\dagger} \varphi_N )_{\mathbf{1^{''}}}    
%\nonumber  \\
+\rho_7 (\varphi_E \varphi_E )_{\mathbf{1^{''}}} 
(\varphi_N^{\dagger} \varphi_N )_{\mathbf{1^{'}}}+H.c.,
\label{VphiNE}
\end{align}
\begin{align}
V_6(\varphi_E, \varphi_N) =& \kappa_1 (\varphi_E \varphi_E )_{\mathbf{3_s}} \varphi_N+\kappa_2 (\varphi_E \varphi_E )_{\mathbf{3_a}} \varphi_N  
%\nonumber  \\
+\kappa_3 (\varphi_N^{\dagger} \varphi_N )_{\mathbf{3_s}} \varphi_E+\kappa_4 (\varphi_N^{\dagger} \varphi_N )_{\mathbf{3_a}} \varphi_E+H.c.,
\end{align}
\end{widetext}
and Yukawa interactions involving $\varphi_E$, $\varphi_N$ and charged leptons, 
\begin{widetext}
\begin{align}
\label{lagran}
-\mathcal{L}_{Y}^f = &~\lambda _{e}^f (\overline{l}_L \phi_{h}) \tilde{e}_R \frac{\varphi_N}{\Lambda}
+ \lambda _{\mu}^f \left( \overline{l}_L \phi_{h} \right)^{''} \tilde{\mu}_R \frac{\varphi_N}{\Lambda} 
%\nonumber \\
%&
+\lambda _{\tau}^f \left( \overline{l}_L \phi_{h} \right)^{'} \tilde{\tau}_R \frac{\varphi_N}{\Lambda} 
+ g_N^f \left( \overline{N^c} N \right) \varphi_E+H.c.,
\end{align} 
\end{widetext}
because such interactions would destroy too much the charged lepton mass structure, 
which is already described relatively well by the SM (see more below).\\
 
Let us denote the VEV's of these scalar fields $\xi$, $\xi^{'}$, $\xi^{''}$, 
$\varphi_E:=(\phi_1,\phi_2,\phi_3)$ and  
$\varphi_N:=(\varphi_1,\varphi_2,\varphi_3)$
as follows 
$$
\langle \xi \rangle = \sigma_a, ~  \langle \xi^{'} \rangle = \sigma_b, 
~  \langle \xi^{''} \rangle = \sigma_c,$$ 
\begin{equation}
\langle \phi_h \rangle =v_h,~\langle \varphi_E \rangle = \left(v_1, v_2, v_3 \right), ~ \langle \varphi_N \rangle 
= \left(u_1,u_2,u_3 \right).
\label{svev}
\end{equation}
To get a VEV of $\varphi_E=(\phi_1,\phi_2,\phi_3)$ imposes 
an extremum condition on the potential $V$,
\begin{equation}
\left. \frac{\partial V}{\partial \phi_i}\right 
\vert _{\langle \phi_i \rangle=v_i}=0,~~~~ (i=1,2,3),
\label{phi-min}
\end{equation}
\begin{widetext}
leading to the equation system of $v_i$
\begin{equation}
\begin{cases}
2(\alpha_1+\alpha_3^{'})v_1^3+(\alpha_2-\alpha_3^{'})(v_2^3+v_3^3)+4(\alpha_1+\alpha_2)v_1v_2v_3+\alpha_{6_r}v_1\sigma_b \sigma_c=0,  \\
2(\alpha_1+\alpha_2)v_1^2v_3+3(\alpha_2-\alpha_3^{'})v_1v_2^2+(4 \alpha_1+\alpha_2+3\alpha_3^{'})v_2v_3^2+\alpha_{6_r}v_3 \sigma_b \sigma_c=0,  \\
2(\alpha_1+\alpha_2)v_1^2v_2+3(\alpha_2-\alpha_3^{'})v_1v_3^2+(4 \alpha_1+\alpha_2+3\alpha_3^{'})v_2^2v_3+\alpha_{6_r}v_2 \sigma_b \sigma_c=0,  
\end{cases}
\end{equation}
\end{widetext}
where
\begin{equation}
\alpha_3^{'}=\frac{4 \alpha_3}{9},~~~\alpha_{6_r}=\frac{1}{2}(\alpha_6+\alpha_6^{\ast}).
\end{equation}
In principle, this equation system has several solutions but we choose the one satisfying 
the equality  
\begin{equation}
v_1^2= v^2 =\frac{-\alpha_{6_r} \sigma_b \sigma_c}{2 (\alpha_1+\alpha_3^{'})} ,~~~v_2=v_3=0,
\label{vev-phiH}
\end{equation}
in order to get, as shown below, a diagonalized mass matrix of the charged leptons. 
We note that if the fields 
$\xi^{'}$ and $\xi^{''}$ are excluded from the model, the VEV in \eqref{vev-phiH} becomes a 
trivial one, $v_1=v_2=v_3=0$, leading, as seen in \eqref{Ml}, to 
massless charged leptons.\\

Next, for the VEV of $\varphi_N=(\varphi_1,\varphi_2,\varphi_3)$ we have 
the equations 
\begin{widetext}
\begin{equation}
\begin{cases}
\lambda_0 u_1+2(\lambda_1+\lambda_3^{'})u_1^3+(2\lambda_2-\lambda_3^{'}+\lambda_5^{'})(u_2^3+u_3^3)
+2(2\lambda_1+4\lambda_2-\lambda_5^{'})u_1u_2u_3+\beta_2 u_3\\+\beta_3 u_2=0, \\[2mm]
\lambda_0 u_3+2(\lambda_1+2\lambda_2+\lambda_5^{'})u_1^2u_3+(6\lambda_2-3\lambda_3^{'}-\lambda_5^{'})u_1u_2^2
+ (4\lambda_1+2\lambda_2+3\lambda_3^{'}-\lambda_5^{'})u_2u_3^2  \\
+\beta_2 u_2+\beta_3 u_1=0,  \\[2mm]
\lambda_0 u_2+2(\lambda_1+2\lambda_2)u_1^2u_2+(6\lambda_2-3\lambda_3^{'}-\lambda_5^{'})u_1u_3^2
+ (4\lambda_1+2\lambda_2+3\lambda_3^{'}+\lambda_5^{'})u_2^2u_3+\beta_2u_1\\+\beta_3u_3=0, 
\end{cases}
\label{vev-phiN-eq}
\end{equation}
\end{widetext}
where
\begin{align}
&\lambda_0=\mu^2+\gamma_1 \sigma_a^2+\gamma v_h^2,~~\lambda_3^{'}=\frac{4 \lambda_3}{9}, ~~\lambda_5^{'}=\frac{4 \lambda_5}{9},  \\
&\beta_2 = \gamma_2 \sigma_c^2,~~\beta_3 = \gamma_3 \sigma_b^2.
\end{align} 
This equation system has a special solution with 
\begin{equation}
u_1^2=u_2^2=u_3^2 =-\frac{\lambda_0+\beta_2+\beta_3}{6(\lambda_1+2\lambda_2)}\equiv u^2 
\label{1u}
\end{equation}
and another solution with 
\begin{equation}
u_1 \neq u_2 \neq u_3 \neq u_1,
\label{3u} 
\end{equation}
which, however, has a too long expression in order to be written down here (in fact, 
we do not need its explicit analytical expression but below numerical calculations 
will be done). As we will see later, the solution \eqref{1u} leads to a TBM model, 
while the solution \eqref{3u} leads to a non-TBM model. 
%\\

\subsubsection{Lepton sector}

~~~ Basing on the $A_4\times Z_3\times Z_4$ flavour symmetry we can construct the following 
Yukawa terms of the effective 
Lagrangian for the lepton sector of the present model:\\
\begin{widetext}
\begin{align}
\label{lagran}
-\mathcal{L}_{Y} = &~\lambda _{e} (\overline{l}_L \phi_{h}) \tilde{e}_R \frac{\varphi_E}{\Lambda}+ 
\lambda _{\mu} \left( \overline{l}_L \phi_{h} \right)^{''} \tilde{\mu}_R \frac{\varphi_E}{\Lambda}
+\lambda _{\tau} \left( \overline{l}_L \phi_{h} \right)^{'} \tilde{\tau}_R \frac{\varphi_E}{\Lambda}
+ \lambda_D \overline{\ell}_L \tilde{\phi}_h N  \nonumber \\
&+ g_N \left( \overline{N^c} N \right) \varphi_N + g_{\xi} \left(\overline{N^c} N \right)_{\mathbf{1}} 
\xi 
+ H.c..
\end{align}
\end{widetext}
From this Lagrangian we get the following mass matrix of the charged leptons,
\begin{equation}
\begin{matrix}
M_l=v_h \left(
\begin{array}{lcr}
\frac{\lambda_e v_1}{\Lambda} & \frac{\lambda_{\mu} v_2}{\Lambda} & \frac{\lambda_{\tau} v_3}{\Lambda} \\[2mm]
\frac{\lambda_e v_3}{\Lambda} & \frac{\lambda_{\mu} v_1}{\Lambda} & \frac{\lambda_{\tau} v_2}{\Lambda} \\[2mm]
\frac{\lambda_e v_2}{\Lambda} & \frac{\lambda_{\mu} v_3}{\Lambda} & \frac{\lambda_{\tau} v_1}{\Lambda}.
\end{array}
\right).
\end{matrix}
\label{Ml}
\end{equation}
~\\
As explained above, we choosed the VEV alignment 
\eqref{vev-phiH}, 
\begin{equation}
\langle \varphi_E \rangle = (v,0,0).
\end{equation}
This choice of the VEV of $\varphi_E$ breaks the symmetry $A_4$ down to its subgroup $G_S$ 
\cite{Altarelli:2009kr}.
The corresponding charged lepton mass matrix automatically has a diagonal form 
\begin{equation}
\begin{matrix}
M_l=  \left(
\begin{array}{ccc}
y_ev_h & 0       & 0\\
  0 & y_{\mu}v_h & 0\\
  0 & 0       & y_{\tau}v_h
\end{array}
\right)
\end{matrix},
\end{equation}
where
\begin{equation}
y_e = \frac{\lambda_e v}{\Lambda},~~y_{\mu}=\frac{\lambda_{\mu} v}{\Lambda},
~~y_{\tau}=\frac{\lambda_{\tau} v}{\Lambda}.
\end{equation}
It is obvious that $v$ must be non-zero ($v\neq 0$), otherwise, the charged 
leptons are massless (this case happens when $\xi^{'}$ and $\xi^{''}$ are 
absent or they develop no VEV).\\

For the neutrino mass matrix, the Majorana part $M_N$ and the Dirac 
part $M_D$ are respectively 
\begin{align}
\label{MM}
M_N& =
\begin{matrix}
\left(
\begin{array}{ccc}
2b_1+d & -b_3  & -b_2\\
-b_3 & 2b_2 & -b_1+d \\
-b_2 & -b_1+d & 2b_3
\end{array}
\right),
\end{matrix}
\end{align}
and 
\begin{equation}
\label{m_dir}
M_D =  \lambda_D v_h \left(
\begin{array}{ccc}
1  &  0  &  0 \\
0 & 0 & 1 \\
0  &  1 & 0
\end{array}
\right),
\end{equation}
where 
\begin{align}
d =2 g_{\xi} \sigma_a, ~
b_1 =\frac{2}{3} g_N u_1, ~b_2 =\frac{2}{3} g_N u_2, ~b_3 = \frac{2}{3} g_N u_3.  
\label{bi} 
\end{align}
From the seesaw mechanism 
\cite{Bilen,Carlo,Mohapatra:1998rq,Minkowski:1977sc,GellMann:1980vs,Mohapatra:1979ia,
Schechter:1980gr,Schechter:1981cv}, we get a neutrino mass matrix 
of the form 
\begin{equation}
\label{M_neu}
M_{\nu}=-M_D^TM_N^{-1}M_D.
\end{equation}
As the scale of $M_M$ is very large but not fixed yet (however, the relative scale 
\eqref{M_neu} is important) we can work, for a further 
convenience, in a scale where $M_D$ is normalized to 1, that is, 
$(\lambda_Dv_h)^2 \sim 1$.
It is not difficult to see that for the VEV alignment $u_1=u_2=u_3=u$ in \eqref{1u}, 
that is,  
$b_1=b_2=b_3\equiv b$, 
the matrix \eqref{M_neu} has the form 
\begin{widetext}
\begin{equation}
M_{\nu 0}= \frac{1}{{\cal D}_0}\left(
\begin{array}{ccc}
3b^2+2b d-d^2 & -3b^2+b d & -3b^2+b d \\
-3b^2+b d & 3 b^2+2 b d & 3b^2- b d-d^2 \\
-3b^2+b d & 3b^2- b d-d^2 & 3 b^2+2 b d
\end{array}
\right) \equiv \frac{1}{{\cal D}_0} M'_0,
\label{M_v0}
\end{equation}
\end{widetext}
where ${\cal D}_0\equiv \det(M^0_{N})$, taking the value
\begin{align}
{\cal D}_0 = 9b^2d-d^3,
\end{align}
is the determinant  ${\cal D}\equiv \det(M_{N})$ of the matrix $M_N$ for $u_1=u_2=u_3$.
It can be checked that the mass matrix $M_{\nu 0}$, as noted above, can be diagonalized by the 
TBM matrix (up to a phase factor) 
\begin{equation}
\label{utbm}
U_{tbm} =
\begin{matrix}
\left(
\begin{array}{ccc}
\sqrt{\frac{2}{3}} & \sqrt{\frac{1}{3}} & 0 \\
-\sqrt{\frac{1}{6}} & \sqrt{\frac{1}{3}} & -\sqrt{\frac{1}{2}} \\
-\sqrt{\frac{1}{6}} & \sqrt{\frac{1}{3}} & \sqrt{\frac{1}{2}}
\end{array}
\right).
\end{matrix}
\end{equation}

For the VEV alignment $u_1 \neq u_2 \neq u_3 \neq u_1$ of $\varphi_N$ in \eqref{3u} 
the neutrino mass has a general form  
\begin{equation}
M_{\nu}=-M_D^TM_N^{-1}M_D = 
\left(
\begin{array}{ccc}
A & B & C \\
B & E & D \\
C & D & F
\end{array}
\right),
\label{Mn}
\end{equation} 
where $A$, $B$, $C$, $D$, $E$ and $F$ in general are complex numbers but here we do 
not need their explicit expressions. 
One of the key problems of a neutrino mass and mixing model is to diagonalize the 
corresponding neutrino mass matrix. Customarily, instead of \eqref{Mn}, the matrix 
\begin{equation}
\label{M2m}
\MM_{\nu} \equiv M_{\nu}M_{\nu}^{\dagger}
\end{equation}
must be diagonalized. Let $U_{pmns}$ be the matrix diagonalizing the matrix 
\eqref{M2m},    
\begin{equation}
\label{M2diag}
\text{diag}(\MM_{\nu})=U_{pmns}^{\dagger}\MM_{\nu} U_{pmns}.
\end{equation}
Here, $U_{pmns}$ is a mixing matrix, which may differ from the PMNS matrix, denoted as 
$U_{PMNS}$, by a phase factor. It is a difficult task to find a realistic 
(phenomenological) model to realize $U_{pmns}$, i.e., $U_{PMNS}$. To solve this problem, 
different methods and 
tricks have been used. Since, as discussed earlier, $U_{pmns}$ slightly differs from the TBM 
form \eqref{utbm} we will follow a perturbation approach. 
This approach allows us to find a theoretical mixing matrix, 
say $U$, which must be compared with the empirical PMNS matrix. 
\section{Neutrino masses and mixing}

~~~ The standard (three-) neutrino mixing matrix, the PMNS matrix, has the canonical 
form (upto a diagonal phase matrix to be specified below)
\begin{widetext}
\begin{equation}
\label{gupmns}
U_{pmns}=
\left(
\begin{array}{ccc}
 c_{12}c_{13} & s_{12}c_{13} & s_{13} e^{-i \delta} \\
 -c_{23}s_{12}-s_{13}s_{23}c_{12} e^{i \delta} & c_{23}c_{12}-s_{13}s_{23}s_{12} e^{i \delta} & s_{23}c_{13} \\
 s_{23}s_{12}-s_{13}c_{23}c_{12} e^{i \delta} & -s_{23}c_{12}-s_{13}c_{23}s_{12} e^{i \delta} & c_{23}c_{13}
\end{array}
\right),
\end{equation}
\end{widetext}
where
$s_{ij}=\sin \theta_{ij}$, $c_{ij}=\cos\theta_{ij}$ with  
$\theta_{ij}  \in [0,\pi/2]$ being mixing angles, and  
$\delta\equiv \delta_{CP} \in [0,2\pi]$ being the Dirac CPV phase.
In a TBM model (for which $s_{13}=0$, $s_{23}^2= \frac{1}{2}$, $s_{12}^2=\frac{1}{3}$) 
this matrix $U_{pmns}$ becomes the matrix $U_{tbm}$ in \eqref{utbm}.
Here we work with the choice $s_{23}=-\sqrt{\frac{1}{2}}$, $s_{12}=\sqrt{\frac{1}{3}}$ 
but another choice, for example, $s_{23}=\sqrt{\frac{1}{2}}$, $s_{12}=\sqrt{\frac{1}{3}}$,  
can be made. \\

The current experimental data 
($\theta_{13} \approx 9^{\circ}$, $\theta_{23} \approx 42^{\circ}$, $\theta_{12} \approx 33^{\circ}$) 
\cite{Capozzi:2013csa} shows that the matrix $U_{pmns}$ can be obtained from  $U_{tbm}$ by a small 
correction as seen from their difference 
\begin{equation}
\left(
\begin{array}{ccc}
 0.006 & -0.029 & 0.153 e^{-i \delta} \\
 -0.008+0.084 e^{i \delta} & 0.047\, +0.056 e^{i \delta} & 0.054 \\
 0.041-0.095 e^{i \delta} & -0.027\, -0.064 e^{i \delta} & 0.034 \\
\end{array}
\right).
\label{deltaU}
\end{equation}
Therefore, we can consider $U_{pmns}$ as a perturbative development around $U_{tbm}$. This requirement 
will impose a restriction on the construction of a model, in particular, on its parameters.
Working in the basis of the diagonalized charged lepton mass matrix (i.e., in the basis $l=e, \mu, \tau$) 
and with a neutrino mixing matrix treated as a small deviation from the TBM form, one can write a 
perturbative expansion of $\MM_{\nu}$ around a non-perturbative TBM mass matrix $\MM_0$, which can be 
diagonalized (cf. \cite{Dinh:2015tna,Brahmachari:2012cq}), 
\begin{equation}
U_{TBM}^{\dagger}\MM_0 U_{TBM}=\text{diag}(|m_{01}|^2, |m_{02}|^2, |m_{03}|^2),
\label{vmv}
\end{equation}
by the matrix 
\begin{equation}
U_{TBM} =
\begin{matrix}
\left(
\begin{array}{ccc}
\sqrt{\frac{2}{3}} & \sqrt{\frac{1}{3}} & 0 \\
-\sqrt{\frac{1}{6}} & \sqrt{\frac{1}{3}} & -\sqrt{\frac{1}{2}} \\
-\sqrt{\frac{1}{6}} & \sqrt{\frac{1}{3}} & \sqrt{\frac{1}{2}}
\end{array}
\right)
\end{matrix}\times P_0
\sim \left( \vert 1^0 \rangle, \vert 2^0 \rangle, \vert 3^0 \rangle \right),
\label{Utbm}
\end{equation}
where $m_{0i}$, $i=1,2,3$, are non-perturbative masses, 
and 
\begin{equation}
P_0=\mbox{diag} \left(e^{i {\alpha_{01}\over 2}}, e^{i {\alpha_{02}\over 2}}, 1 \right),
\end{equation}
with $\alpha_{01}$ and $\alpha_{02}$ being Majorana phases. 
We note that $U_{tbm}$ given in \eqref{utbm} differs from $U_{TBM}$ \eqref{Utbm}, 
used frequently in the literature, by the factor $P_0$. 
 Thus, $M_{\nu}$ in \eqref{Mn} 
can be written 
as 
\begin{equation}
M_{\nu} = M_0 +{\cal V},
\end{equation}
with 
\begin{equation}
M_0= \frac{M'_0}{ {\cal D}}, ~~ {\cal D}= \det(M_N),
\end{equation}
where $M'_0$ is defined in  \eqref{M_v0} and ${\cal V}$ is a small matrix to be specified below. 
At the first 
order of perturbation the matrix $\MM$ is developed around $\MM_0$ as follows  
\begin{equation}
\MM_{\nu} = \MM_0 + (M_0^\dagger {\cal V} + {\cal V}^\dagger M_0).
\end{equation}
Thus the squared masses $|m_i|^2$ obtained by a diagonalization of $\MM$ represent a perturbative 
shift 
\begin{equation}
|m_i|^2=|m_{0i}|^2+\delta |m_i|^2
\label{pertu-mass}
\end{equation}
from the non-perturbative squared masses $|m_{0i}|^2$, where $m_{0i}$ now have the form 
\begin{equation}
m_{01}= {(3b-d)d\over {\cal D}}, ~ m_{02}= {9b^2-d^2\over {\cal D}}, 
~ m_{03}= {(3b+d)d\over {\cal D}}.
\end{equation}

Since a homogeneous VEV alignment $\langle \varphi_N \rangle= (u,u,u)$ 
such as that in \eqref{1u} leads 
to a TBM mixing but the experiment tells us a mixing slightly deviating from the 
TBM one, we must consider an inhomogeneous VEV alignment \eqref{3u} to deviate from 
a homogeneous alignment with an appropriate amount, that is     
\begin{equation}
(u_1,u_2,u_3)  = (u_1,u_1+\epsilon_2,u_1+\epsilon_3),
\label{u123}
\end{equation}
where $(0, \epsilon_2, \epsilon_3)$ is an appropriate shift of 
$\langle \varphi_N \rangle$ from the level $(u_1,u_1,u_1)$. It can be shown 
that it is enough this shift to obey the condition $\epsilon_2, \epsilon_3 \ll {\cal D}/g_N$, 
if not stronger, $\epsilon_2, \epsilon_3 \ll 1$.
The latter can be satisfied 
if $\lambda_1$, $\lambda_2$, $\lambda'_3$ and $\lambda'_5$ are chosen to have the same 
order of magnitude but much bigger than that of $\lambda_{0}$, i.e., 
\begin{equation}
\lambda_{0} \ll \lambda_1\approx \lambda_2\approx  \lambda'_3\approx \lambda'_5 \equiv \lambda
\label{lambda}
\end{equation}
as well as $\beta_2$ and $\beta_3$ are chosen to be at the same order of magnitude but much smaller 
than that of $\lambda $, i.e., 
\begin{equation}
\beta_2\approx \beta_3 \ll \lambda.
\label{beta}
\end{equation} 
It is observed from \eqref{bi} that an alignment $(u_1,u_2,u_3)$ is proportional to an alignment 
$(b_1,b_2,b_3)$, therefore, a homogeneous alignment $(b,b,b)$ corresponds to a TBM mixing. That 
means that a realistic alignment $(b_1,b_2,b_3)$ must deviate from a homogeneous alignment 
by only a small amount:
\begin{equation}
(b_1,b_2,b_3)  = (b_1,b_1+e_2,b_1+e_3),
\label{b123}
\end{equation}
where $e_2, e_3 \ll {\cal D}$ 
(see \eqref{e23D} below for a numerical inllustration).  
Taking into account  \eqref{lambda} -- \eqref{b123} we get 
\begin{widetext}
\begin{equation}
{\cal V}=\frac{1}{{\cal D}} \left(
\begin{array}{ccc}
4b(e_2+e_3)  & -de_3+b(4e_2+e_3)  & -d e_2+b(e_2+4e_3)  \\
-de_3+b(4e_2+e_3) & 4 b e_2+2 de_2-2be_3  & b (e_2+e_3)   \\
-d e_2+b(e_2+4e_3) & b (e_2+e_3)  & 4be_3+2de_3-2be_2
\end{array}
\right).
\label{Vcal}
\end{equation}
\end{widetext}

Now a perturbation expansion is made around the TBM state \eqref{Utbm}. 
Here, we will follow the perturbative approach described in 
\cite{Sakurai:2011zz}.
Using the perturbation decomposition 
\begin{equation}
|n\rangle=|n^0\rangle+\sum_{k\neq n}a_{kn}|k^0\rangle +...,
\end{equation}
with $\vert n^0 \rangle$ defined in \eqref{Utbm} and 
\begin{equation}
a_{kn}=(|m_{0n}|^2-|m_{0k}|^2)^{-1}V_{kn},~~
V_{kn}=\langle k^0|M_0^\dagger {\cal V} + {\cal V}^\dagger M_0|n^0\rangle,
\label{per-exp}
\end{equation}
one can diagonalize the matrix $\MM_{\nu}$, 
\begin{equation}
{U}^{\dagger}\MM_{\nu}U=\mbox{diag}\left(|m_1|^2,|m_2|^2,|m_3|^2 \right),
\end{equation}
by the matrix 
\begin{widetext}
\begin{equation}
\label{upmns}
U= U_{TBM}+\Delta U =
\begin{matrix}
\left(
\begin{array}{llr}
~~\sqrt{\frac{2}{3}}+\Delta U_{11} & \sqrt{\frac{1}{3}} +\Delta U_{12} & \Delta U_{13}\\
-\sqrt{\frac{1}{6}}+\Delta U_{21} & \sqrt{\frac{1}{3}}+\Delta U_{22}  &- \sqrt{\frac{1}{2}}+\Delta U_{23} \\
-\sqrt{\frac{1}{6}}+\Delta U_{31} & \sqrt{\frac{1}{3}}+\Delta U_{32}  & \sqrt{\frac{1}{2}}+\Delta U_{33}
\end{array}
\right)\times P_0,
\end{matrix}
\end{equation}
\end{widetext}
representing a perturbative expansion from $U_{TBM}$ in \eqref{utbm}, 
where (upto the first perturbation order)
\newpage 
\begin{widetext}
\begin{equation}
\Delta U_{11}=  \sqrt{ \frac{1}{3}} X^{\ast}, ~~
\Delta U_{12}= - \sqrt{\frac{2}{3}}X,~~ \Delta U_{13}= - \sqrt{\frac{2}{3}}Y-\sqrt{\frac{1}{3}}Z,  
\nonumber 
\end{equation}
\begin{equation}
\Delta U_{21}= \sqrt{\frac{1}{3}}X^{\ast}-\sqrt{\frac{1}{2}}Y^{\ast}, ~~
\Delta U_{22}= \sqrt{\frac{1}{6}}X-\sqrt{\frac{1}{2}}Z^{\ast},~~
\Delta U_{23}= \sqrt{\frac{1}{6}}Y-\sqrt{\frac{1}{3}}Z,
\nonumber 
\end{equation}
\begin{equation}
\Delta U_{31}= \sqrt{\frac{1}{3}}X^{\ast}+ \sqrt{\frac{1}{2}}Y^{\ast}, ~~
\Delta U_{32}= \sqrt{\frac{1}{6}}X+\sqrt{\frac{1}{2}}Z^{\ast},~~
\Delta U_{33}= \sqrt{\frac{1}{6}}Y-\sqrt{\frac{1}{3}}Z,
\end{equation}
\end{widetext}
and
\begin{equation}
X = -a_{12}, ~~Y=-a_{13},~~Z=-a_{23}.
\end{equation}
We note that the parameters $a_{ij}$ defined in \eqref{per-exp} and appearing in $\Delta U$, 
are determined from the elements of the matrix $\cal{V}$ in \eqref{Vcal} derived under the 
condition \eqref{b123} leading to imposing constraints \eqref{lambda} and \eqref{beta} on 
the model parameters.\\

To check the model how it works, let us make a numerical analysis. It is enough (and for 
simplicity) to assume the parameters $g_N, d,  \lambda_0, \lambda_1$ to be real. Under 
this assumption the equation system \eqref{vev-phiN-eq} has 27 solutions 
$\left( u_1, u_2, u_3 \right)$ belonging to the following four types:

\begin{align}
\text{Type-1:}~~& \left(0,~ 0,~ 0 \right),~~\text{i.e.},~ u_1=u_2=u_3=0,  \\ 
\text{Type-2:}~~& \left(u,~ 0,~ 0 \right), ~~ u\neq 0, \\
\text{Type-3:}~~& \left(u,u,u\right), ~~ u\neq 0, \\
\text{Type-4:}~~& \left(u_1,u_2,u_3 \right); ~ u_1\neq u_2\neq u_3\neq u_1, 
~ u_i\neq 0.
\label{sol-type}
\end{align}
It is observed that the solutions of type-1, type-2 and type-3 do not lead to the PMNS 
mixing as desired (where only the type-3 solutions give the TBM mixing), therefore, 
they are excluded from our consideration and only the solutions of type-4 remains at choice.\\

One of the type-4 solutions having the form  
\begin{widetext}
\begin{equation}
( u_1, u_2, u_3)=
\left(-(0.14+0.28 i)\sqrt{\frac{\lambda_0}{\lambda}}, 
-(0.019-0.32 i)\sqrt{\frac{\lambda_0}{\lambda}}, 
-(0.17-0.26 i)\sqrt{\frac{\lambda_0}{\lambda}} \right) 
\end{equation}
\end{widetext}
gives a result consistent with the current experimental data (see below). It follows 
\begin{widetext}
\begin{equation}
(b_1, b_2, b_3)=\left(-(0.14+0.28 i)K, -(0.019-0.32 i)K, -(0.17-0.26 i)K \right), ~~K = {2\over 3}g_N \sqrt{\frac{\lambda_0}{\lambda}}.
\label{bK}
\end{equation}
\end{widetext}
The neutrino masses \eqref{pertu-mass} now get the form \cite{Sakurai:2011zz} 
\begin{align}
m_1^2=m_{01}^2+V_{11},~ 
%\nonumber \\
m_2^2=m_{02}^2+V_{22}, ~ 
%\nonumber\\
m_3^2=m_{03}^2+V_{33}, 
\label{mi2}
\end{align}
where $V_{ii}$ are given in \eqref{per-exp}, namely, 

$$V_{ii}=\langle i^0 \vert M_0^{\dagger}{\cal V}+{\cal V}^{\dagger}M_0 \vert i^0 \rangle, ~~i=1, 2, 3.$$
Using the experimental data for the squared mass differences 
$\Delta m_{21}^2$ and $\Delta m_{32}^2$ (see Table \ref{Synopsis} below):  
\begin{align}
\Delta m_{21}^2 = m_2^2-m_1^2 = 7.54 \cdot 10^{-5}, \nonumber  \\
\Delta m_{31}^2 = m_3^2-m_1^2 = 2.47 \cdot 10^{-3}.
\label{deltam2}
\end{align}
we can find $K$ in \eqref{bK} and $d$ in \eqref{MM}. Here, for a demonstration, 
we work with a normal neutrino mass ordering, but the case with an inverse 
neutrino mass ordering is similar. Since the equations \eqref{deltam2}  
are non-linear in $K$ and in $d$, they may have more than one solutions in $K$ 
and in $d$. Below, as an illustration, we will expose one of the numerical solutions, 
\begin{equation}
K = 1.74+0.05 i, ~~~d = -9.01,
\label{k2}
\end{equation}
giving  
\begin{equation}
{e_2\over {\cal D}} = 0.0003+0.0015 i, ~{e_3\over {\cal D}} = -0.0001+0.0014 i.
\label{e23D}
\end{equation}
and 
\begin{align}
&X = 0.326+0.034 i, ~~
Y = -0.007+0.003 i,  \nonumber  \\
&Z = -0.082+0.251 i.
\end{align}
The latter values of $X$, $Y$ and $Z$ provide 
\begin{equation}
U_{13}=0.053-0.148 i.
\label{u13}
\end{equation}
It is not difficult to find all other elements of $U$ and ${\cal V}$ which we do not 
expose here to save the paper's length. Further, using \eqref{k2} in \eqref{deltam2} 
we obtain absolute neutrino masses
\begin{equation}
m_1 = 0.1109 ~\text{eV},~m_2=0.1114 ~\text{eV},~m_3=0.1217 ~\text{eV}.
\end{equation}
This result is consistent with the current experimental data \cite{Agashe:2014kda} 
and it means that our model and method work quite well.\\

From \eqref{u13}, as $U_{13}= s_{13}e^{-i\delta}$, we obtain
 $s_{13}\approx 0.157$ 
(or $\theta_{13}\approx 9.03 ^{\circ}$) and $\delta \approx 1.39 \pi$. 
The latter value of $s_{13}$ is very close to the experimental data shown in \eqref{deltaU}. 
Interestingly, the Dirac CPV phase, $\delta_{CP}\equiv \delta$, obtained here, surprisingly 
(but hopefully not just accidentally) 
coincides with its global fit given in \cite{Capozzi:2013csa}. A more detailed analysis on 
$\delta_{CP}$ will be make in the next section.
\section{Dirac CP violation phase and Jarlskog parameter}

~~~ In order to determine all variables in the matrix \eqref{upmns}, or, at least, their relations, 
we must compare this matrix with the experimental one. Denoting the elements of the matrix 
\eqref{upmns} by $U_{ij}$, $i,j=1,2,3$, we get the equation (up to the first perturbation order) 
\begin{equation}
\label{ueq}
2 \left(|U_{21}|^2-|U_{31}|^2 \right) -\left(|U_{22}|^2-|U_{32}|^2 \right)=-2 \sqrt{2} Re(U_{13}).
\end{equation}
Further, comparing $U_{ij}$ in \eqref{ueq} with the corresponding elements of the matrix $U_{PMNS}$ 
given in the "trigonometric" form 
\begin{widetext}
\begin{equation}
\label{gupmns}
U_{PMNS}=
\left(
\begin{array}{ccc}
 c_{12}c_{13} & s_{12}c_{13} & s_{13} e^{-i \delta} \\
 -c_{23}s_{12}-s_{13}s_{23}c_{12} e^{i \delta} & c_{23}c_{12}-s_{13}s_{23}s_{12} e^{i \delta} & s_{23}c_{13} \\
 s_{23}s_{12}-s_{13}c_{23}c_{12} e^{i \delta} & -s_{23}c_{12}-s_{13}c_{23}s_{12} e^{i \delta} & c_{23}c_{13}
\end{array}
\right)\times  P\equiv U_{pmns}\times  P,
\end{equation}
\end{widetext}
where $P$ (which in general is different from $P_0$) is a diagonal matrix of the form
$$P=\mbox{diag}\left(e^{i \frac{\alpha_1}{2}}, e^{i \frac{\alpha_2}{2}},1 \right) $$
with $\alpha_1$ and $\alpha_2$ being Majorana phases, we obtain the following relation between the 
Dirac CPV phase 
$\delta_{CP}\equiv\delta$ and the neutrino mixing angles $\theta_{ij}$,
\begin{widetext}
\begin{align}
\label{re2}
&\left(c_{23}^2 -s_{23}^2 \right) \left(2 s_{12}^2-c_{12}^2 \right)+12s_{13}s_{23}c_{23}s_{12}c_{12} 
\cos \delta 
= -2 \sqrt{2} s_{13} \cos \delta,
\end{align}
\end{widetext}
neglecting $\mathcal{O}(\lambda^2)$ terms and higher order perturbation terms. Solving this equation for 
$\cos\delta$ we get
%%%%%%%%%%%%%%%%%%%%%%%%%%
\begin{equation}
 \cos \delta = \frac{(s_{23}^2-c_{23}^2)(2s_{12}^2-c_{12}^2)}{2 \sqrt{2}(3 
\sqrt{2}s_{23}c_{23}s_{12}c_{12}+1)s_{13}}.
 \label{cdel}
\end{equation}
Below, this equation will be used for $s_{23}<0$ because, as seen in \eqref{utbm} and \eqref{Utbm}, 
its TBM limit ($-\sqrt{1/2}$) is negative, while the small perturbative fluctuation cannot change 
its sign.
Since we work with $\delta\in [0,2\pi]$, if $\delta_0$ is a solution of the  equation \eqref{cdel} 
so is $2\pi-\delta_0$. Having a value of $\delta_{CP}$ we can obtain a value of the Jarlskog parameter $J_{CP}$. 
\begin{widetext}
\begin{center}
\begin{table}[H]
\begin{center}
%\vspace*{2mm}
\begin{tabular}{|l|c|c|c|c|}
\hline
%\vspace{1mm}
\quad\quad\quad~ Parameter & Best fit & $1\sigma$ range & $2\sigma$ range & $3\sigma$ range\\
%&&&& \\
\hline%---------------------------------------------------------------------
$\Delta m_{21}^2/10^{-5}~\mathrm{eV}^2 $ (NO or IO) & 7.54 & 7.32 -- 7.80 & 7.15 -- 8.00 & 6.99 -- 8.18 \\
\hline%---------------------------------------------------------------------
$\sin^2 \theta_{12}/10^{-1}$ (NO or IO) & 3.08 & 2.91 -- 3.25 & 2.75 -- 3.42 & 2.59 -- 3.59 \\
\hline%---------------------------------------------------------------------
$\Delta m_{31}^2/10^{-3}~\mathrm{eV}^2 $ (NO) & 2.47 & 2.41 -- 2.53 & 2.34 -- 2.59 & 2.27 -- 2.65 \\
$|\Delta m_{32}^2|/10^{-3}~\mathrm{eV}^2 $ (IO) & 2.42 & 2.36 -- 2.48 & 2.29 -- 2.55 & 2.23 -- 2.61 \\
\hline%---------------------------------------------------------------------
$\sin^2 \theta_{13}/10^{-2}$ (NO) & 2.34 & 2.15 -- 2.54 & 1.95 -- 2.74 & 1.76 -- 2.95 \\
$\sin^2 \theta_{13}/10^{-2}$ (IO) & 2.40 & 2.18 -- 2.59 & 1.98 -- 2.79 & 1.78 -- 2.98 \\
\hline%---------------------------------------------------------------------
$\sin^2 \theta_{23}/10^{-1}$ (NO) & 4.37 & 4.14 -- 4.70 & 3.93 -- 5.52 & 3.74 -- 6.26 \\
$\sin^2 \theta_{23}/10^{-1}$ (IO) & 4.55 & 4.24 -- 5.94 & 4.00 -- 6.20 & 3.80 -- 6.41\\
\hline
\end{tabular}
\end{center}
\caption{\label{Synopsis} Experimental data for a normal ordering (NO) and an inverse ordering (IO) 
\cite{Capozzi:2013csa,Agashe:2014kda}.}
%}
\end{table}
\end{center}
\end{widetext}

\begin{figure}[h!]
\centering
    \includegraphics[width=0.4\textwidth]{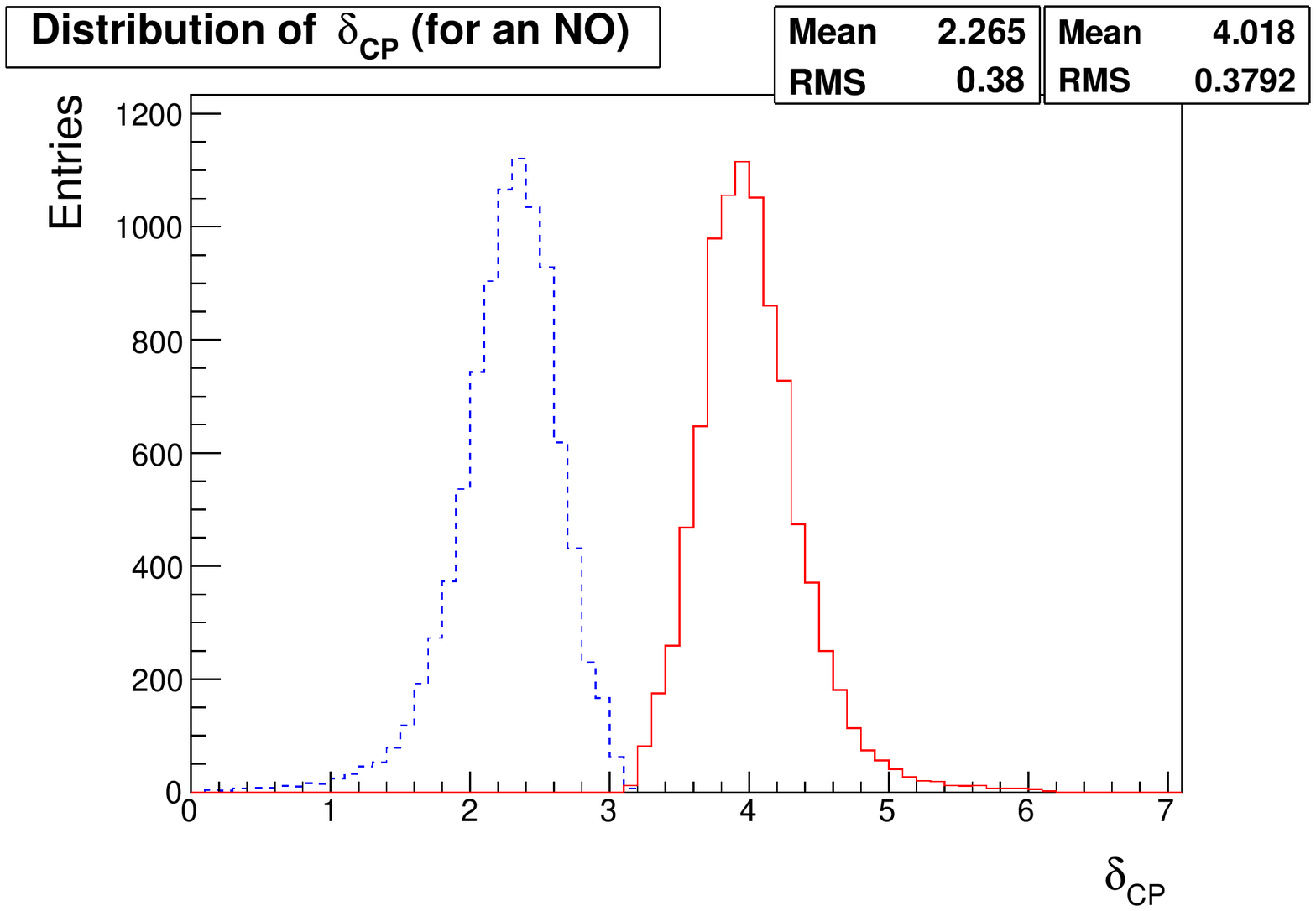} 
    \vspace*{-6mm}
\caption{Distribution of $\delta_{CP}$ in an NO.}
\label{delta}
\end{figure}

\begin{figure}[h!]
\centering
    \includegraphics[width=0.4\textwidth]{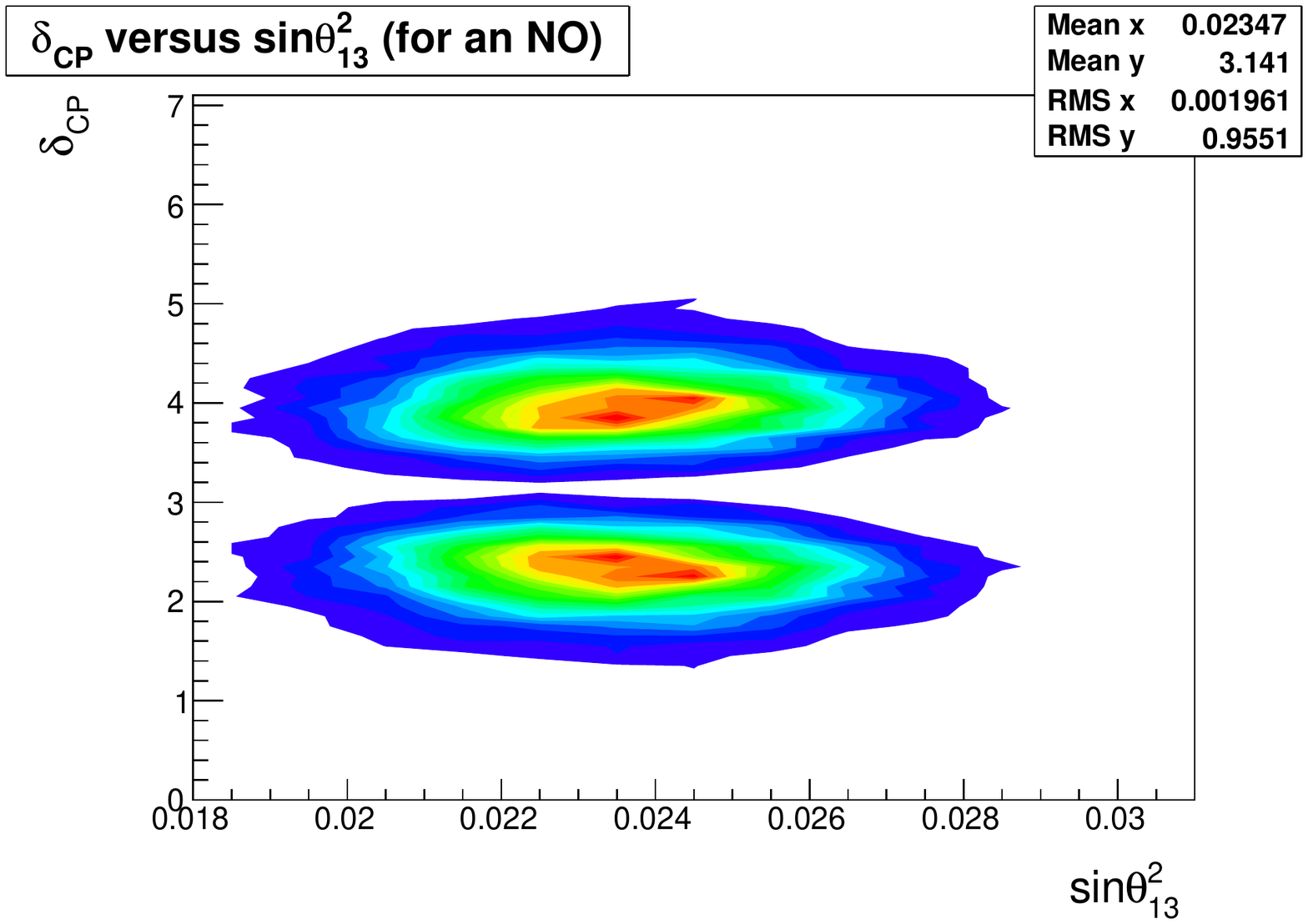}
    \vspace*{-5mm}
    \caption{$\delta_{CP}$ versus $\sin^2\theta_{13}$ in an NO.}
    \label{sindelta}
%\end{center}
\end{figure}

\begin{figure}[h!]
	\centering
	\includegraphics[width=0.4\textwidth]{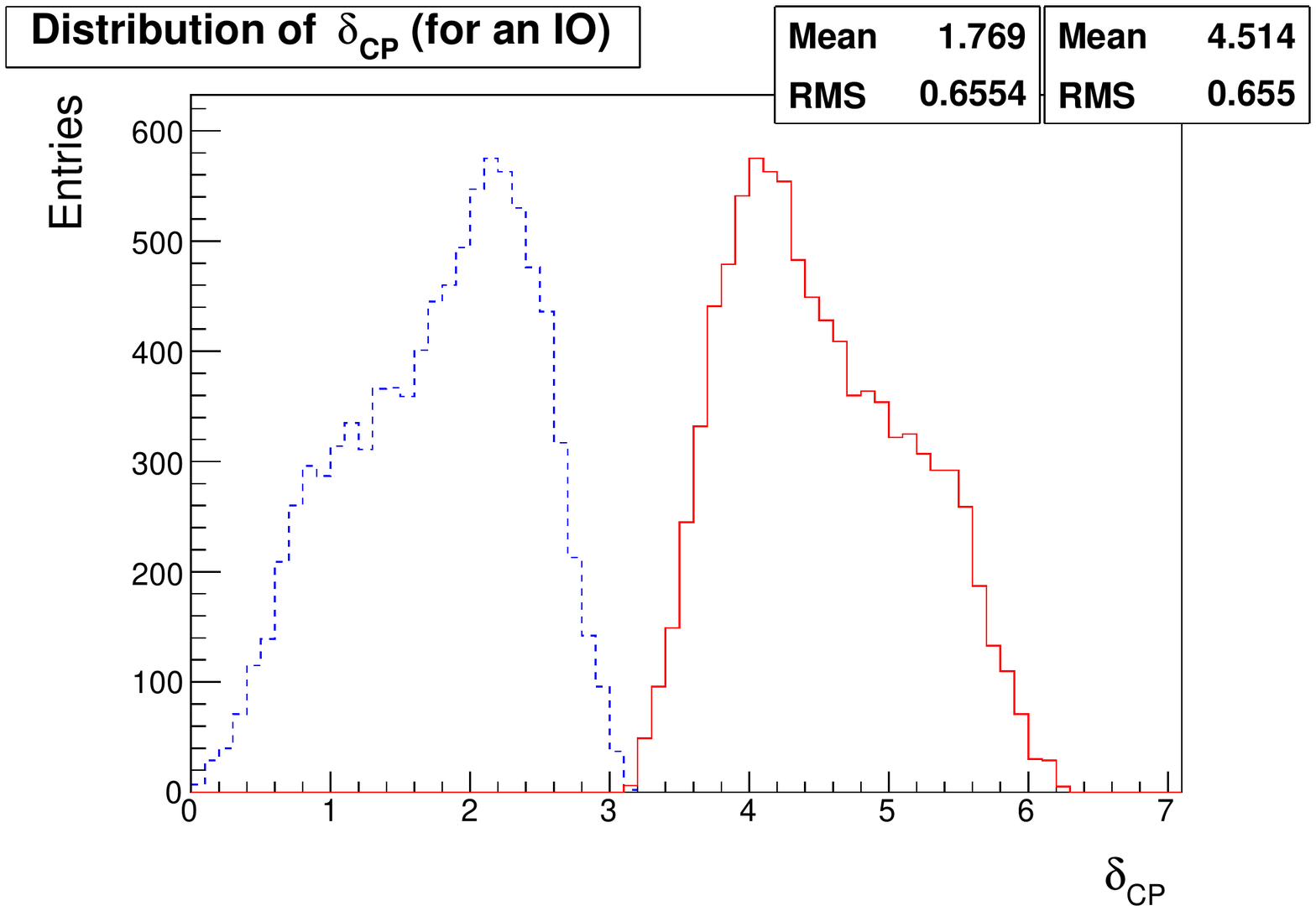} 
	\vspace*{-6mm}
	\caption{Distribution of $\delta_{CP}$ in an IO.}
	\label{ihdelta}
\end{figure}

\begin{figure}[h!]
\vspace*{4mm}
	\centering
	\includegraphics[width=0.4\textwidth]{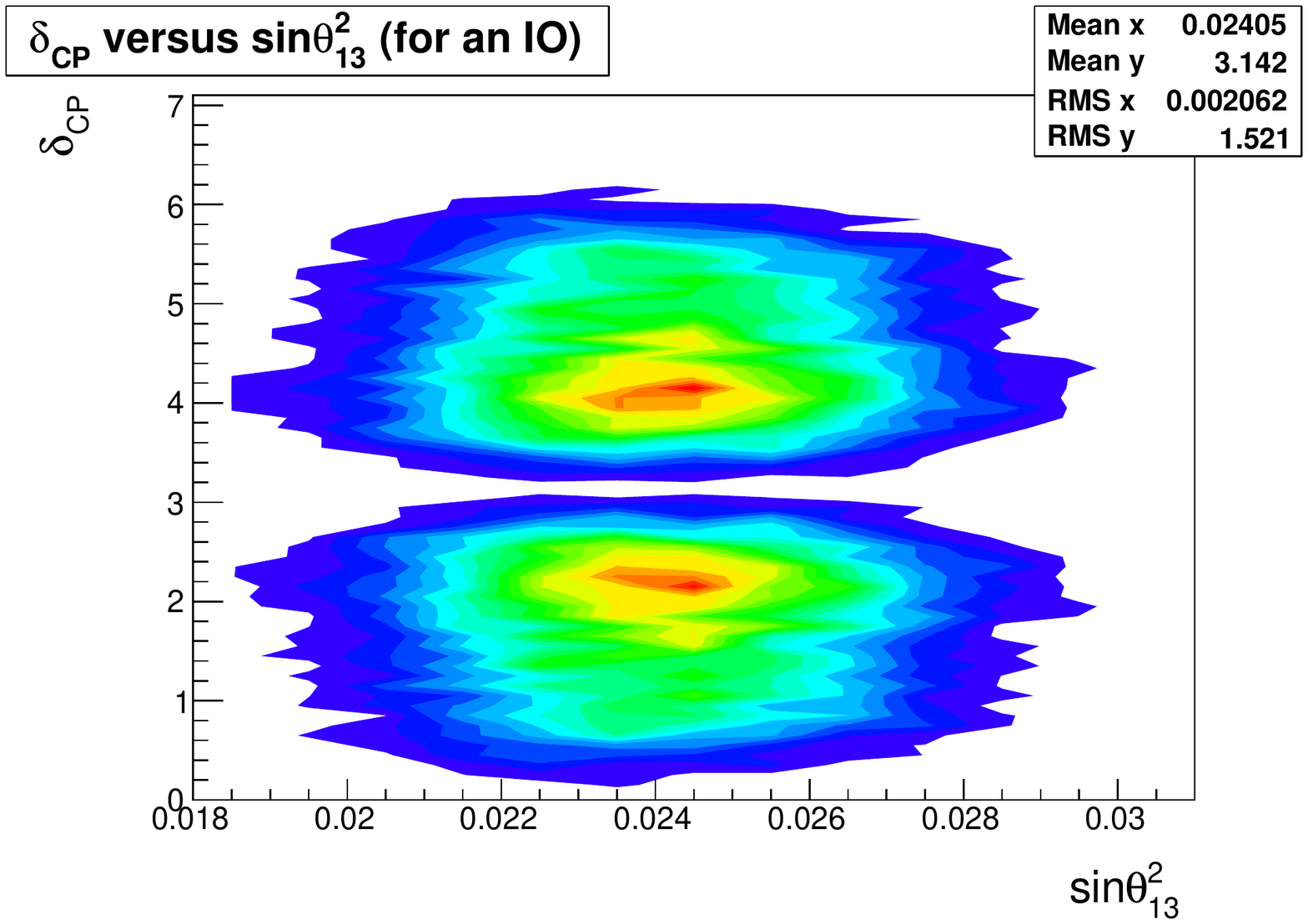} 
	\vspace*{-5mm}
	\caption{$\delta_{CP}$ versus $\sin^2\theta_{13}$ in an IO.}
	\label{sinihdelta}
\end{figure}

Based on the relation \eqref{cdel} and experimental inputs (see Table \ref{Synopsis}), 
$\delta_{CP}$ can be calculated numerically. With using the experimental data of the 
mixing angles within $1\sigma$ around the BFV \cite{Capozzi:2013csa,Agashe:2014kda}, 
the distributions of $\delta_{CP}$ are plotted in Fig. \ref{delta} and Fig. \ref{sindelta} 
for a normal neutrino mass ordering (NO) and in Fig. \ref{ihdelta} and Fig. \ref{sinihdelta} 
for an 
inverse neutrino mass ordering (IO). Here, for each of these distributions, 10000 events 
are created 
and $\delta_{CP}$ is calculated event by event with $s_{ij}$ taken as random values 
generated on the base of a Gaussian distribution having the mean (best fit) value and 
sigmas given in Tab. \ref{Synopsis}. Each of these distributions has two (sub)populations 
corresponding to two solutions of \eqref{cdel}. In Fig. \ref{delta} and Fig. \ref{sindelta}, 
the distributions corresponding to two solutions are distinguished by being plotted in blue 
and red. We see that the solution located in the range $[\pi, 2\pi]$ is nearer the BFV 
(within 1$\sigma$ region). In Fig. \ref{sindelta} and Fig. \ref{sinihdelta}, the three  
1$\sigma$, 2$\sigma$ and 3$\sigma$ regions are colored with different colors (red, green 
and blue, respectively).\\

In the case of an NO, $\delta_{CP}$ has a mean value of $2.265\approx 0.72\pi$ for one of the solutions, 
and a mean value of $4.018\approx 1.28\pi$ for the other solution and its distribution gets maximums at 
$2.35\approx 0.75\pi$ and $3.95\approx 1.26\pi$, respectively. We see that the second solution 
(for both its mean value and the value at its maximal distribution) lies in the $1\sigma$ region from 
the best fit value (BFV) $1.39 \pi$ given in \cite{Capozzi:2013csa,Agashe:2014kda}. \\

In the case of an IO, $\delta_{CP}$ gets a mean value around $1.769\approx 0.56\pi$ (for the first solution), 
and around $4.514\approx 1.44\pi$ (for the second solution). Its distribution reaches maximums at about 
$2.15 \approx 0.68\pi$ and $4.17 \approx 1.33\pi$. Again, the second solution lies within the 
1$\sigma$ region of the BFV $1.31\pi$ given in \cite{Capozzi:2013csa,Agashe:2014kda}. \\

\begin{figure}[h!]
	\centering
	\includegraphics[width=0.4\textwidth]{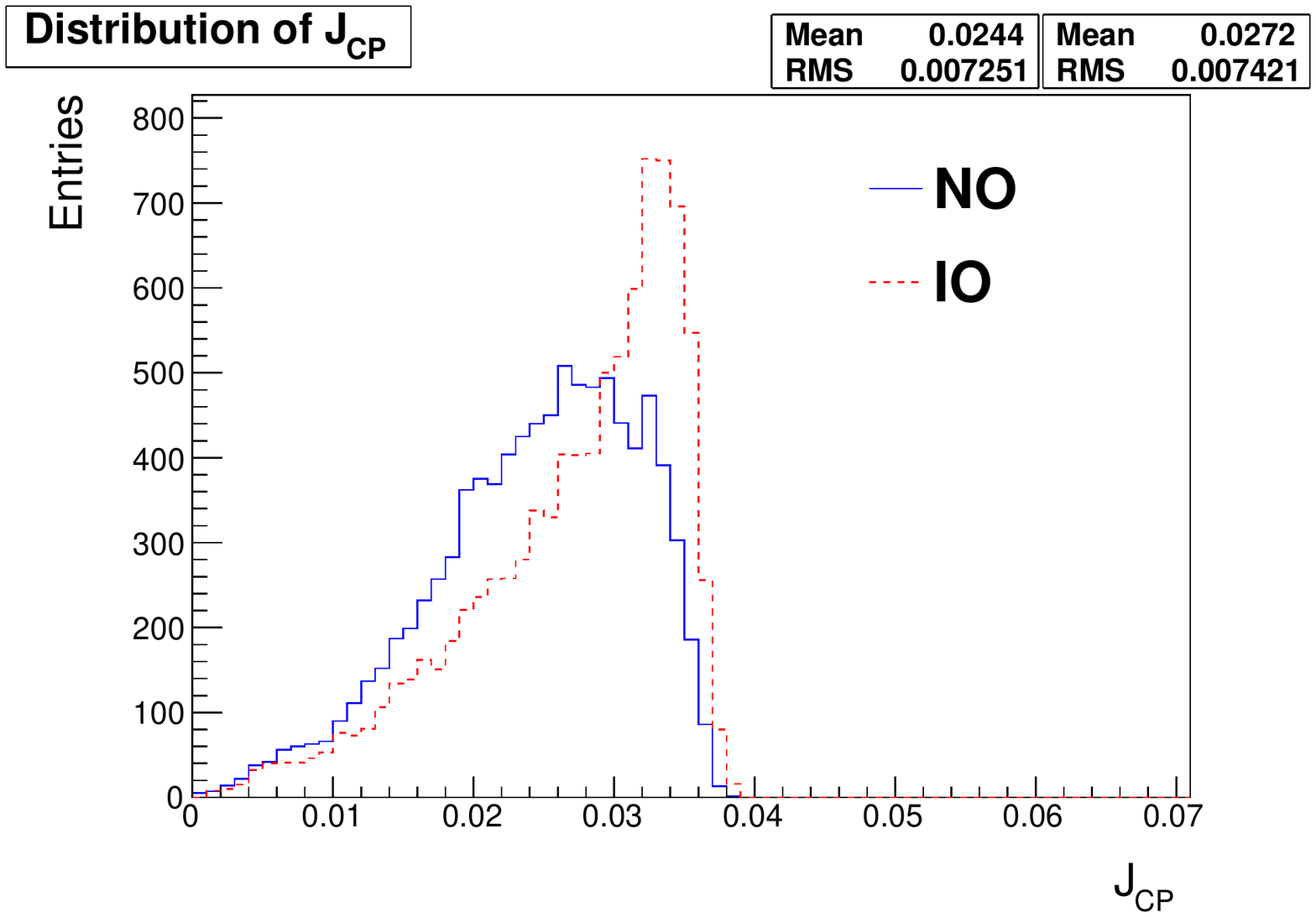} 
	\vspace*{-5mm}
	\caption{Distribution of $J_{CP}$ in an NO and an IO.}
	\label{J2h}
\end{figure}

Having all mixing angles and Dirac CPV phase it is not difficult to determine the Jarlskog parameter 
$J_{CP}\equiv J$. Indeed, using the expression \cite{Bilen}
\begin{equation}
|J_{CP}|=|c_{12}c_{23}c_{13}^2s_{12}s_{23}s_{13}\sin\delta |,
\label{Jcp}
\end{equation}
we obtain $|J_{CP}|\leq 0.038$ and  $|J_{CP}|\leq 0.039$ (rough bounds) for an NO and an IO, respectively 
(see the distribution of $J_{CP}$ in Fig. \ref{J2h}). 
It, upto a sign, has a mean value and 
a maximum at 
\begin{equation}
J_{mean}^{NO}=0.024 ~~\text{and} ~~ J_{max}^{NO}=0.027,
\label{jm-no}
\end{equation}
respectively, for an NO, and 
\begin{equation}
J_{mean}^{IO}=0.027 ~~\text{and} ~~  J_{max}^{IO}=0.033,
\label{jm-io}
\end{equation}
respectively, for an IO. The result obtained here is similar to that obtained in 
\cite{Girardi:2015zva,Girardi:2014faa,Petcov:2014laa,Petcov:2013poa,Chakdar:2014ifa} 
by other methods by other authors. 

To have a better view in comparing the two cases, the NO and the IO, the BFV's of 
$\delta_{CP}$ and $J_{CP}$ for both cases are summarized in Table \ref{dj}. These 
mean values of $\delta_{CP}$ and $J_{CP}$ are closer to the global fits 
than their corresponding values obtained at the BFV's of the mixing angles (by 
inserting the latter in the analytical expressions \eqref{cdel} and \eqref{Jcp} for 
$\delta_{CP}$ and $|J_{CP}|$, respectively).
\vspace*{5mm}
\begin{table}[h!]
	%~\\[4mm]
	\begin{center}
	\begin{tabular}{|l||c|c|}
		\hline
		& Normal ordering & Inverse ordering 
		
		\\  \hline
		\vspace*{-2mm}
		&&\\
		\vspace*{-2mm}
		$\delta_{CP}/\pi$& 1.28 & 1.44\\
		& &
		
		\\  \hline
		\vspace*{-2mm}
		&&\\
		\vspace*{-2mm}
		$|J_{CP}|$ & $0.024$ & $0.027$
		
		\\
		&&
		\\
		\hline 
	\end{tabular}
	\caption{The mean values of $\delta_{CP}$ and $|J_{CP}|$ in an NO and an IO.
%\vspace*{2mm}
	}
	\label{dj}
\end{center}
\end{table}
\vspace*{-5mm}
To avoid any confusion, let us stress that
the mean values of $\delta_{CP}$ and $|J_{CP}|$ do not coincide, in fact and in principle, with 
their values obtained at the BFV's of the mixing angles. It means 
that a value of $\delta_{CP}$ 
or $|J_{CP}|$ obtained at a BFV of the mixing angles should not in any way be identified with 
the mean value of the quantity concerned, although in some case they may be 
close to each other. \\

It is also important to note that the equation \eqref{cdel} is ill-defined in the 3$\sigma$ 
region of the mixing angles. It means that this equation of determination of $\delta_{CP}$ 
restricts the dissipation of the mixing angles (that is, the values scattered too far, in 
the 3$\sigma$ region of distribution, are automatically excluded).

\section*{Conclusions}
\vspace*{1mm}

~~~

Basing on the fact that the observed neutrino mixing differs from a TBM one just slightly,  
we have suggested a non-TBM neutrino mixing model corresponding to this observation. This 
model represents an extended standard model acquiring an additional $A_4\times Z_3\times Z_4$ 
flavour symmetry. 
Besides the SM fields assumed now to have also an $A_4\times Z_3\times Z_4$ symmetry structure 
(see Tab. \ref{a4}), this model contains six additional fields, all are $SU(2)_L$ singlets, 
which are one $A_4$-triplet fermion $N$ (right-handed neutrinos),  
two $A_4$-triplet scalars $\varphi_E$ and $\varphi_N$, and three $A_4$-singlet scalars 
$\xi$, $\xi^{'}$ and $\xi^{''}$. The presence of the fields $\xi^{'}$ and $\xi^{''}$ 
(along with the SM Higgs field $\phi_h$) is very important as it guarantees non-zero masses 
of the charged leptons. To avoid unwanted Lagrangian terms two discrete symmetries $Z_3$ 
and $Z_4$ are also introduced. Then, neutrino masses can be generated via Yukawa couplings 
of neutrinos to all scalars but $\varphi_E$. The corresponding neutrino mass 
matrix is obtained for a general VEV structure 
of the scalar field $\varphi_N$. It is observed that the model in general is a non-TBM model, 
but it becomes a TBM model \cite{Harrison:2002er} under a given circumstance with a specific 
VEV alignment of $\varphi_N$ as in \eqref{1u}. 
Because the current experimentally established 
neutrino mixing represents just a small deviation from a TBM mixing we must build a theoretical 
model to 
satisfy this requirement. The latter puts a restriction on the model, in particular, it imposes 
constraints on its parameters. Therefore, the model constructed can be perturbatively developed 
around a TBM model, and, thus, the perturbative method can be applied to our further analysis.\\

As usually, diagonalizing of a mass matrix is a difficult task. Here, within the above-suggested 
model and via a perturbation approach, the obtained neutrino mass matrix can be diagonalized by 
a matrix $U_{PMNS}$ perturbatively expanded around the tri-bi-maximal matrix $U_{TBM}$.
In this way, a relation, see \eqref{cdel}, between the Dirac CPV 
phase and the mixing angles is established. Based on the experimental values of the mixing 
angles this relation allows us to determine the Dirac CPV phase and the Jarlskog 
invariant in a quite good agreement (within the 1$\sigma$ region of the best fit) with the 
recent experimental data at both the normal- and the inverse neutrino mass 
ordering. These hierarchies are not compatible with each other, hence, only one of them, at 
most, can be realized in the Nature, however, none of them, so far, has been confirmed or 
excluded experimentally. Therefore, we here consider both NO and IO, and have obtained results 
in both cases close to the global fit  \cite{Capozzi:2013csa,Agashe:2014kda}. For an illustration 
checking the model, numerical calculations have been also done and give results which 
are in good agreement with the current experimental data.\\ 

The determination of $\delta_{CP}$ and $J_{CP}$ is often both theoretical and experimental 
difficult problem and it can be used to verify the corresponding theoretical neutrino mixing 
model. 
This paper's method allows us to obtain an explicit $\delta_{CP}$ as a function of the mixing 
angles, thus, $\delta_{CP}$, could be determined experimentally via the mixing angles. This 
function in turn isolates the mixing angles from dissipated values in their distribution, i.e., 
the latter should be excluded. 
Our approach is useful and its application to higher order 
perturbations which may give a better fit is our next consideration. 
Finally, the mass spectrum which can be obtained by diagonalizing the mass matrix \eqref{Mn} 
is a subject of analysis to be done in a separate work. \\

{\bf Note added}: After the submission of this paper we have learned about new results 
\cite{T2K} from T2K which are in quite good agreement with our results, in particular, 
the value of $\theta_{13}\approx 9.03^{\circ}$ obtained by us above 
is very close to that, $\theta_{13}\approx 8.47^{\circ}$, given by T2K 
(and to $\theta_{13}\approx 8.8^{\circ}$ in \cite{Agashe:2014kda}). 

\begin{acknowledgments}
\vspace*{4mm}

This work is supported by Vietnam's National Foundation for Science and Technology Development 
(NAFOSTED) under the grant No 103.03-2012.49.\\

Two of us (N.A.K. and N.T.H.V.) would like to thank Kumar Narain for warm hospitality in the Abdus 
Salam ICTP, Trieste, Italy. N.A.K. would like to thank Wolfgang Lerche and Luis Alvarez-Gaume for 
warm hospitality at CERN, Geneva, Switzerland. The authors also thank Dinh Nguyen Dinh for useful 
discussions.
\end{acknowledgments}

\bibliography{apssamp}

\end{document}